\documentclass[twocolumn,aps,superscriptaddress]{revtex4}
\usepackage[T1]{fontenc}
\usepackage[utf8]{inputenc}
\usepackage{amsmath}
\usepackage{amssymb}
\usepackage{graphicx}
\usepackage{esint}
\usepackage[unicode=true,pdfusetitle,
 bookmarks=false,
 breaklinks=false,pdfborder={0 0 1},backref=false,colorlinks=false]
 {hyperref}
\hypersetup{
 pdfstartview=FitH}

\makeatletter
\@ifundefined{textcolor}{}
{%
 \definecolor{BLACK}{gray}{0}
 \definecolor{WHITE}{gray}{1}
 \definecolor{RED}{rgb}{1,0,0}
 \definecolor{GREEN}{rgb}{0,1,0}
 \definecolor{BLUE}{rgb}{0,0,1}
 \definecolor{CYAN}{cmyk}{1,0,0,0}
 \definecolor{MAGENTA}{cmyk}{0,1,0,0}
 \definecolor{YELLOW}{cmyk}{0,0,1,0}
}

\usepackage[english]{babel}
\usepackage{mciteplus}

\makeatother

\begin{document}

\preprint{This line only printed with preprint option}

\title{Prethermalization and dynamical transition in an isolated trapped
ion spin chain}

\author{Zhe-Xuan Gong}

\email{gzx@umich.edu}

\selectlanguage{english}%

\affiliation{Department of Physics, University of Michigan, Ann Arbor, Michigan
48109, USA}

\affiliation{Center for Quantum Information, IIIS, Tsinghua University, Beijing,
100084, P. R. China}

\author{L. -M. Duan}

\affiliation{Department of Physics, University of Michigan, Ann Arbor, Michigan
48109, USA}

\affiliation{Center for Quantum Information, IIIS, Tsinghua University, Beijing,
100084, P. R. China}
\begin{abstract}
We propose an experimental scheme to observe prethermalization and
dynamical transition in one-dimensional XY spin chain with long range
interaction and inhomogeneous lattice spacing, which can be readily
implemented with the recently developed trapped-ion quantum simulator.
Local physical observables are found to relax to prethermal values
at intermediate time scale, followed by complete relaxation to thermal
values at much longer time. The physical origin of prethermalization
is explained by spotting a non-trivial structure in lower half of
the energy spectrum. The dynamical behavior of the system is shown
to cross difference phases when the interaction range is continuously
tuned, indicating the existence of dynamical phase transition. 
\end{abstract}
\maketitle
The dynamical properties of isolated quantum many-body systems have
been under intense interest in recent years \cite{polkovnikov_colloquium:_2011,cazalilla_focus_2010}.
On the theory side, the research has been centered on whether and
how an isolated quantum system approaches thermal equilibrium. While
certain observables are found to relax to equilibrium in some large
systems \cite{deutsch_quantum_1991,srednicki_chaos_1994,rigol_thermalization_2008,rigol_relaxation_2007,reimann_foundation_2008},
it remains unclear on what conditions and time scale equilibration
occurs in generic systems \cite{masanes_complexity_2013,brandao_convergence_2012,reimann_equilibration_2012,gong_comment_2011}.
On the experimental side, recent progress with cold atoms \cite{kinoshita_quantum_2006,trotzky_probing_2012,gring_relaxation_2012}
and trapped ions \cite{friedenauer_simulating_2008,kim_quantum_2010,islam_onset_2011,britton_engineered_2012,islam_emergence_2013,schneider_experimental_2012}
has made it possible to simulate well controlled simple models, such
as one-dimensional (1D) Bose gas and transverse field Ising model.
These quantum systems can be well isolated from the environmental
bath and have long coherence time, while their physical properties
can be measured at individual atomic level, providing an unprecedented
opportunity for studying non-equilibrium dynamics in closed interacting
systems.

A particularly intriguing phenomenon in this context is called \emph{prethermalization}
\cite{berges_prethermalization_2004}, which has been shown to emerge
in various theoretical setups \cite{kollar_generalized_2011,barnett_prethermalization_2011,worm_relaxation_2012},
and experimentally observed in cold atomic gas \cite{gring_relaxation_2012}.
The emergence of prethermalization is characterized by establishment
of quasi-stationary state at intermediate time scale, and followed
by relaxation to stationary state at much longer time scale (thermalization).
Physical origin of prethermalization, however, is still elusive, and
is primarily speculated to be related to quasi-integrability of the
model \cite{barnett_prethermalization_2011,gring_relaxation_2012}. 

In this paper, we propose a new experimental scheme for observing
and studying prethermalization and related dynamical transition in
a XY spin model, which can be implemented with the current trapped-ion
quantum simulator \cite{islam_emergence_2013}. Our model features
long range spin-spin interaction with inhomogeneous lattice spacing,
and unlike many other systems, the prethermalization can occur already
for as few as a dozen of spins, allowing for its observation in current
experimental systems. The prethermalization shown up in this system
has a quite different mechanism and we find that a non-trivial structure
in the energy spectrum resulting from long-range interaction and inhomogeneous
lattice is responsible for the occurrence of prethermalization in
our model. In addition, by tuning the range of interaction with an
experimental knob, we find the dynamical behavior of system exhibits
three different phases: thermalization only, prethermalization followed
by thermalization, and prethermalization only. The transition between
different phases becomes sharper and sharper with increased number
of spins, hinting the existence of dynamical phase transitions \cite{heyl_dynamical_2012}.

\emph{Model and its dynamic}s Our spin model is based on the experimental
system of a chain of ions confined in a linear Paul trap (Fig. \ref{fig1}).
Through proper configuration of Raman beams, the optical dipole force
can generate an effective transverse field Ising model \cite{kim_entanglement_2009,kim_quantum_2010,islam_onset_2011}:

\begin{figure}
\includegraphics[width=0.4\textwidth]{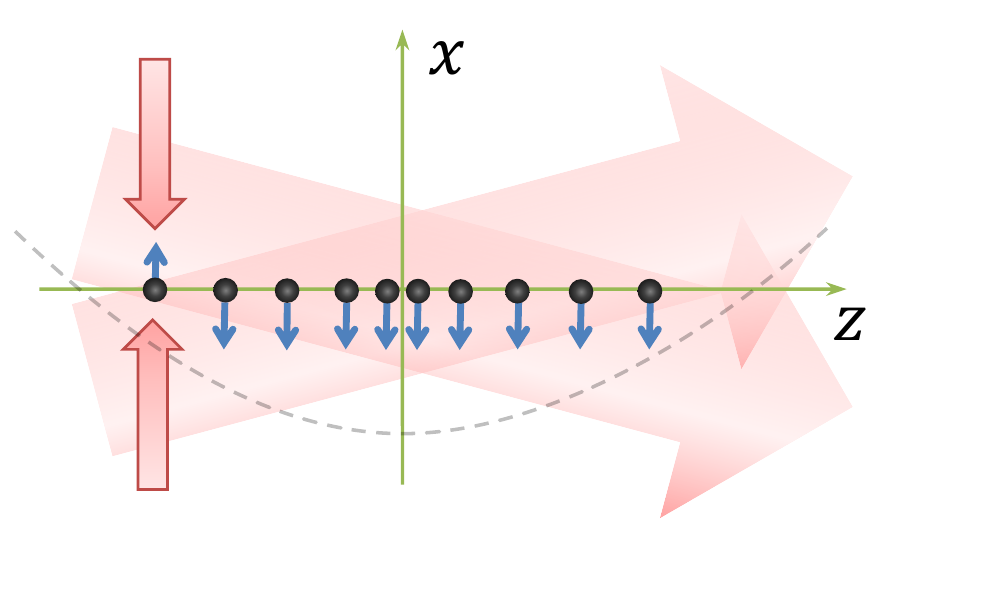}

\caption{\label{fig1}Schematic of the proposed experimental setup: A chain
of $N$ ions are trapped along the $z$ direction in a 1D harmonic
linear Paul trap. The global Raman beams generate spin-dependent force
along $x$ direction, resulting in effective Ising-type interaction.
To induce dynamics, a focused laser beam is applied on one end of
the ion chain to selectively flip only the first spin. }
\end{figure}

\begin{equation}
H=\sum_{i<j}^{N}J_{i,j}\sigma_{i}^{x}\sigma_{j}^{x}+B\sum_{i=1}^{N}\sigma_{i}^{z},\label{TFIM}
\end{equation}
where $\sigma_{i}$ is the spin-1/2 Pauli matrix for the $i^{th}$
ion qubit. The interaction coefficient $J_{ij}$ in Eq. (1) is given
by 
\[
J_{i,j}=\Omega^{2}\sum_{m=1}^{N}\frac{\eta_{i,m}\eta_{j,m}\omega_{m}}{\mu^{2}-\omega_{m}^{2}},
\]
where $\mu$ is the Raman beatnote frequency, $\Omega$ is the effective
Rabi frequency, which is assumed to be uniform on all the ions, $\{\omega_{m}\}$
are the phonon mode frequencies of ions in $x$ direction, and $\eta_{i,m}$
are the Lamb-Dicke parameters measuring the coupling between the $i$th
ion and the $m$th phonon mode. We are interested in the region where
$B\gg\max\{J_{ij}\}$. In this limit, the $\sigma_{i}^{+}\sigma_{j}^{+}$
and $\sigma_{i}^{-}\sigma_{j}^{-}$ terms in Eq. \ref{TFIM} will
be energetically forbidden, and we end up with the $XY$ Hamiltonian:
\begin{equation}
H\approx H_{XY}=\sum_{i<j}2J_{i,j}(\sigma_{i}^{+}\sigma_{j}^{-}+h.c.)+B\sum_{i}\sigma_{i}^{z}\label{XY}
\end{equation}
A unique feature of the Hamiltonian (\ref{TFIM} \& \ref{XY}) realized
with the ion system is that the interaction characterized by $J_{ij}$
is long-ranged and the range of interaction can be readily tuned by
changing the beatnote frequency $\mu$. In particular, in a range
of $\mu$, $J_{ij}$ can be roughly approximated by an power-law decay
with $J_{ij}\sim|i-j|^{-\alpha}$, where $\alpha$ varies from $0$
to $3$ when we tune $\mu$\cite{islam_emergence_2013}. In our following
analysis, for a given $\mu$, we fit the coefficient $J_{ij}$ with
$J_{ij}\sim|i-j|^{-\alpha}$ and use the fitting parameter $\alpha$
as an indicator of the range of iteration.

To study dynamics of the model Hamiltonian (2), we first initialize
all the spins through optical pumping to the spin down state with
$\sigma_{i}^{z}=-1$, which is an eigenstate of $H_{XY}$ and hence
stationary. We then use a focused laser beam to flip the first spin
(left end ion) to $\sigma_{i}^{z}=1$ (see Fig. \ref{fig1}). The
starting state $|\psi(0)\rangle=|\uparrow\downarrow\downarrow\cdots\downarrow\rangle$
is no longer an eigenstate of $H_{XY}$ and subject to dynamics with
$|\psi(t)\rangle=e^{iH_{XY}t/\hbar}|\psi(0)\rangle$. We consider
time evolution of local observables $\langle\sigma_{i}^{z}(t)\}$
and their correlations $\langle\sigma_{i}^{z}(t)\sigma_{j}^{z}(t)\rangle$
which can be directly measured in experiments. For convenience of
description of the dynamics, we introduce the operator 
\[
C=\sum_{i=1}^{N}f_{i}\frac{\sigma_{i}^{z}+1}{2}
\]
where the coefficient $f_{i}\equiv(i-\frac{N+1}{2})/(N-1)$ is equally
distributed between $[-1,1]$ from $i=1$ to $i=N$. The expectation
value of $C$ varies between $[-1,1]$ and physically measures the
relative position of the spin excitation. It's easy to check that
$\langle\psi(0)|C|\psi(0)\rangle=-1,$meaning the spin excitation
is at the left edge of the chain. For any state with spatial inversion
symmetry around the center of the chain, $\langle C\rangle=0$.

\begin{figure}
\includegraphics[width=0.25\textwidth]{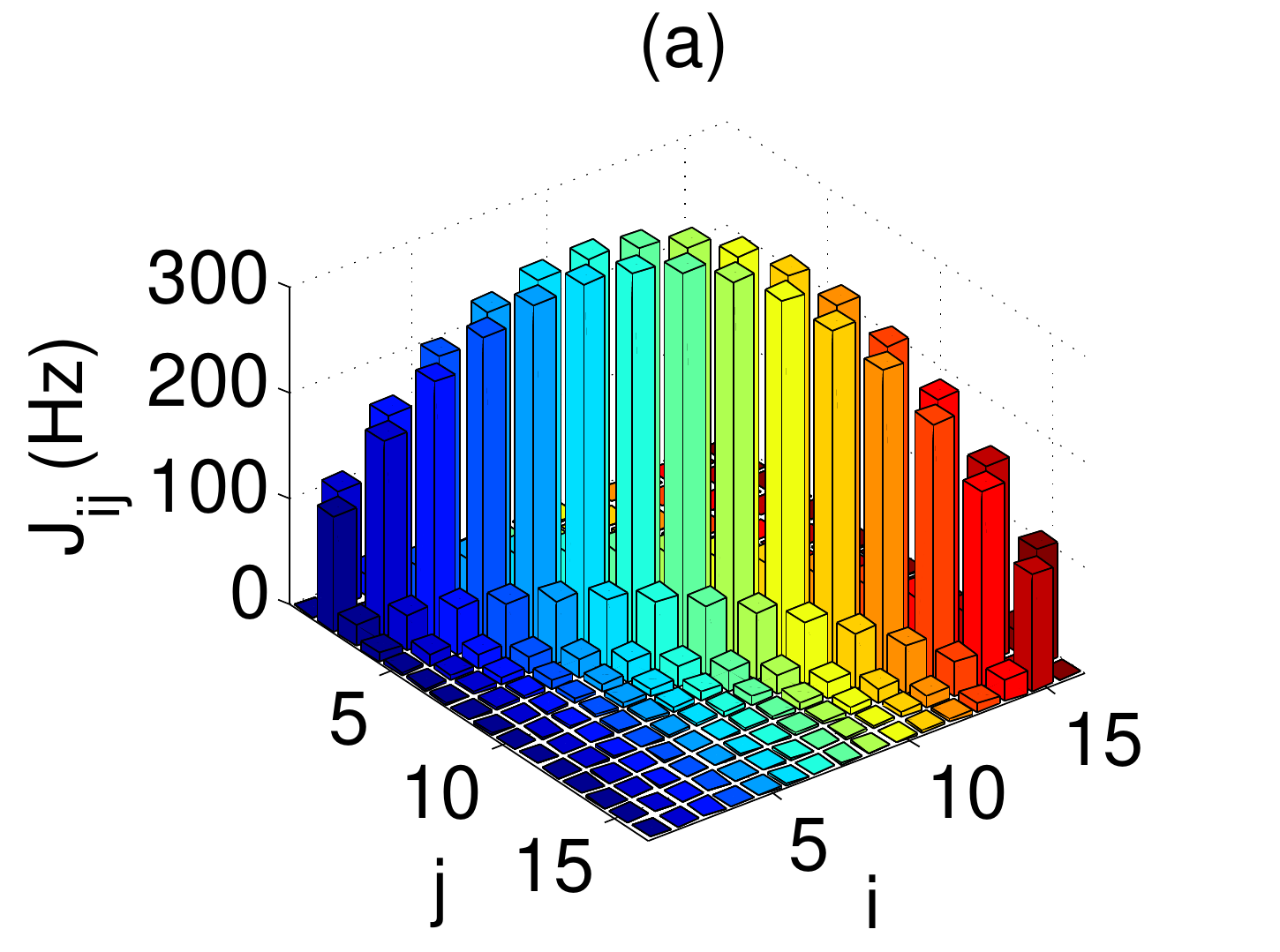}\includegraphics[width=0.25\textwidth]{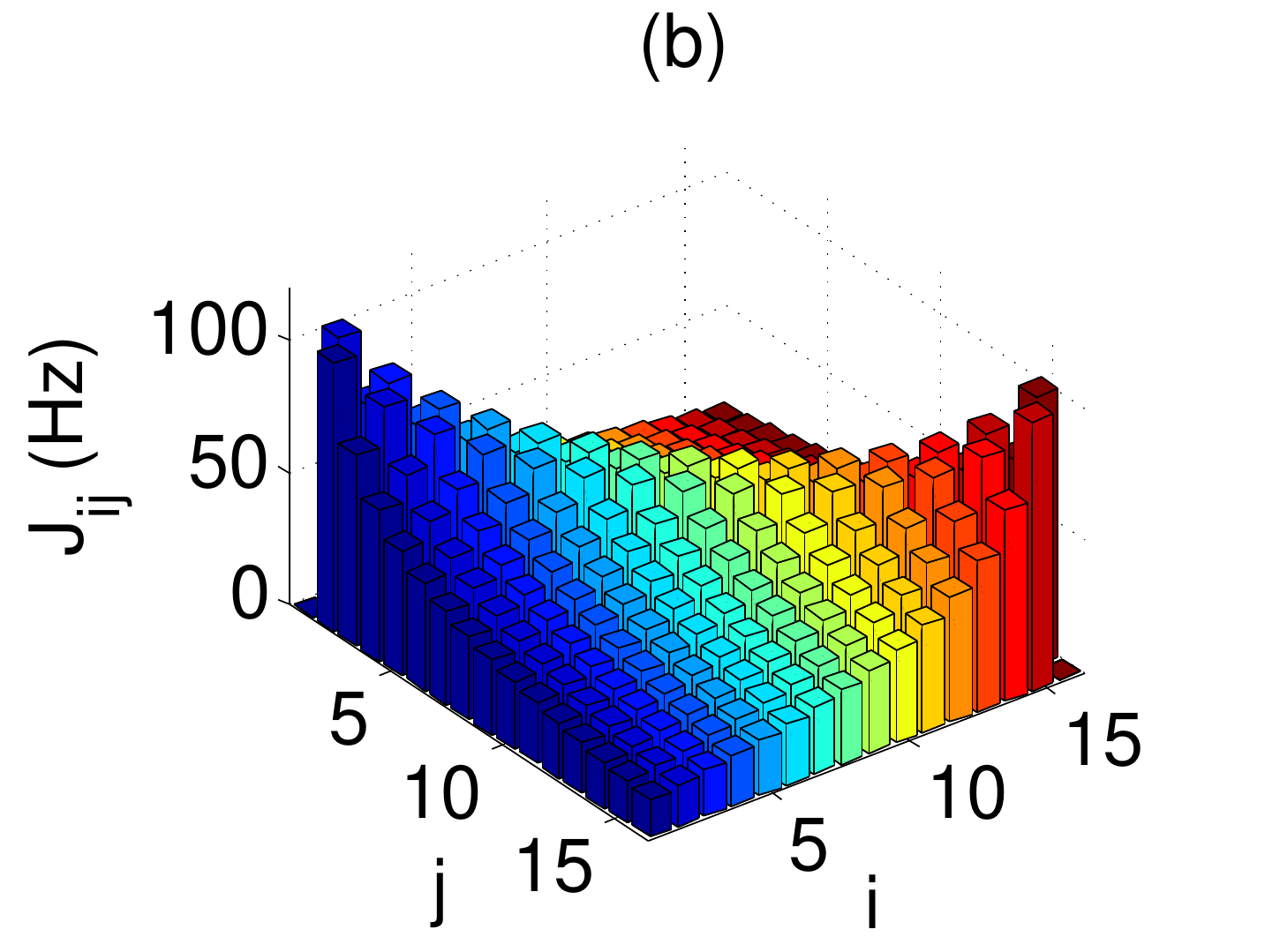}

\caption{\label{fig2}(a) Distribution of $J_{ij}$ for short range interaction
with the beatnote frequency set at $\mu=5.2$ MHz, the trap frequency
in $z$ direction $\omega_{z}=100$ KHz, and $\eta_{x}\Omega=40$
KHz. The corresponding fitting parameter $\alpha\approx2.6$ in this
case. (b) Distribution of $J_{ij}$ for long range interaction with
$\mu=5.02$ MHz, $\omega_{z}=600$KHz, $\eta_{x}\Omega=3.9$KHz, and
the corresponding fitting parameter $\alpha\approx0.52$. In both
cases we have trap frequency in $x$ direction $\omega_{x}=5$ MHz
and average interaction strength $J_{0}=\sum_{i\ne j}J_{ij}/N^{2}=20$
Hz.}
\end{figure}

\emph{Prethermalization and dynamical transition}: To find out the
dynamical behavior, we first perform numerical calculation with an
$N=16$ ion chain, which corresponds to the size of the current experimental
platform for ion quantum simulator \cite{islam_emergence_2013}. As
shown in Fig. \ref{fig2}, we pick up two parameter settings with
the corresponding fitting parameter $\alpha\approx2.6$ and $\alpha\approx0.52$,
which represent respectively short-range and long-range interaction.
The distributions of the exact coupling coefficients $J_{ij}$ are
shown in Fig. \ref{fig2} for these two cases.

\begin{figure}
\includegraphics[width=0.25\textwidth]{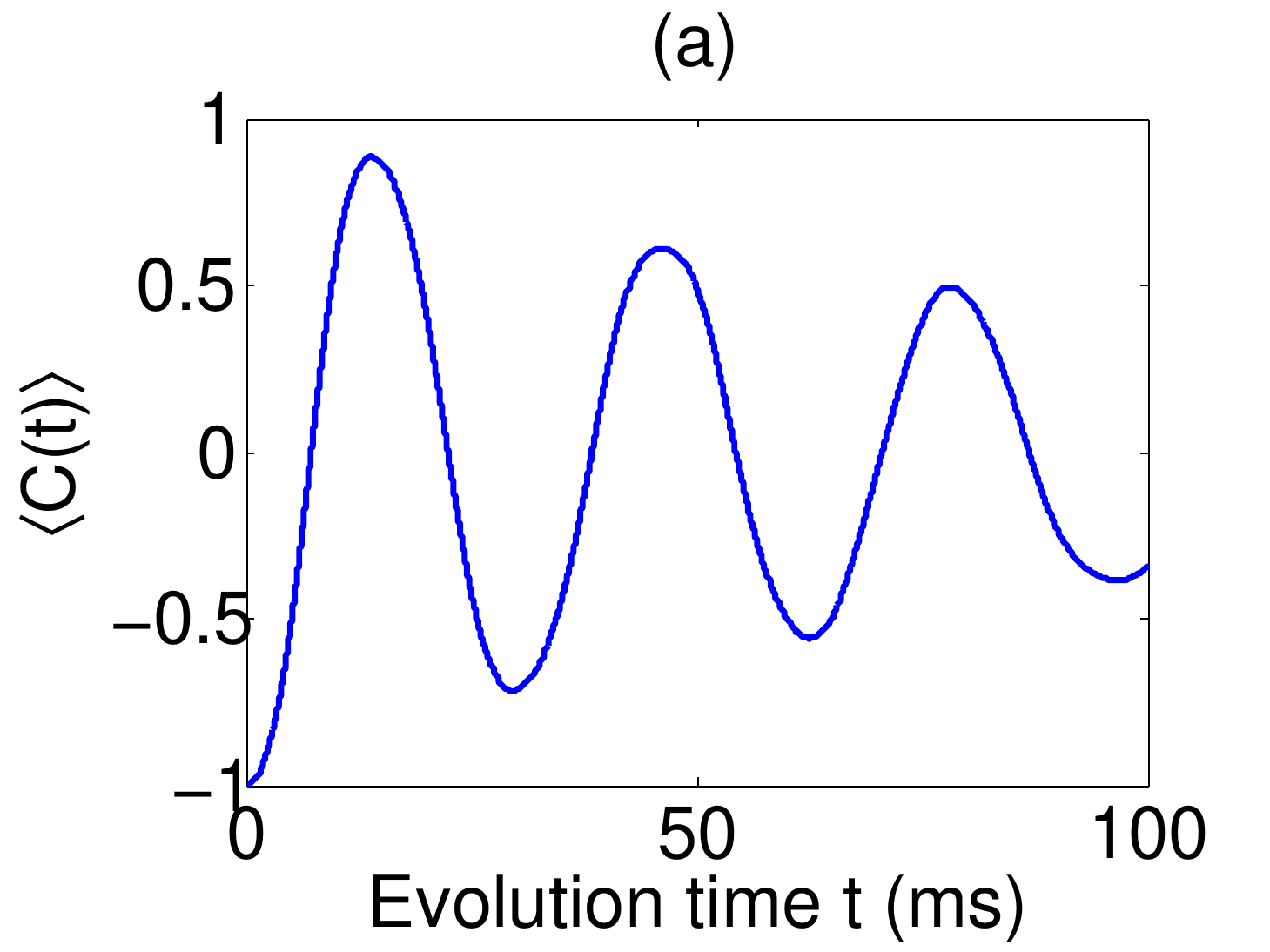}\includegraphics[width=0.25\textwidth]{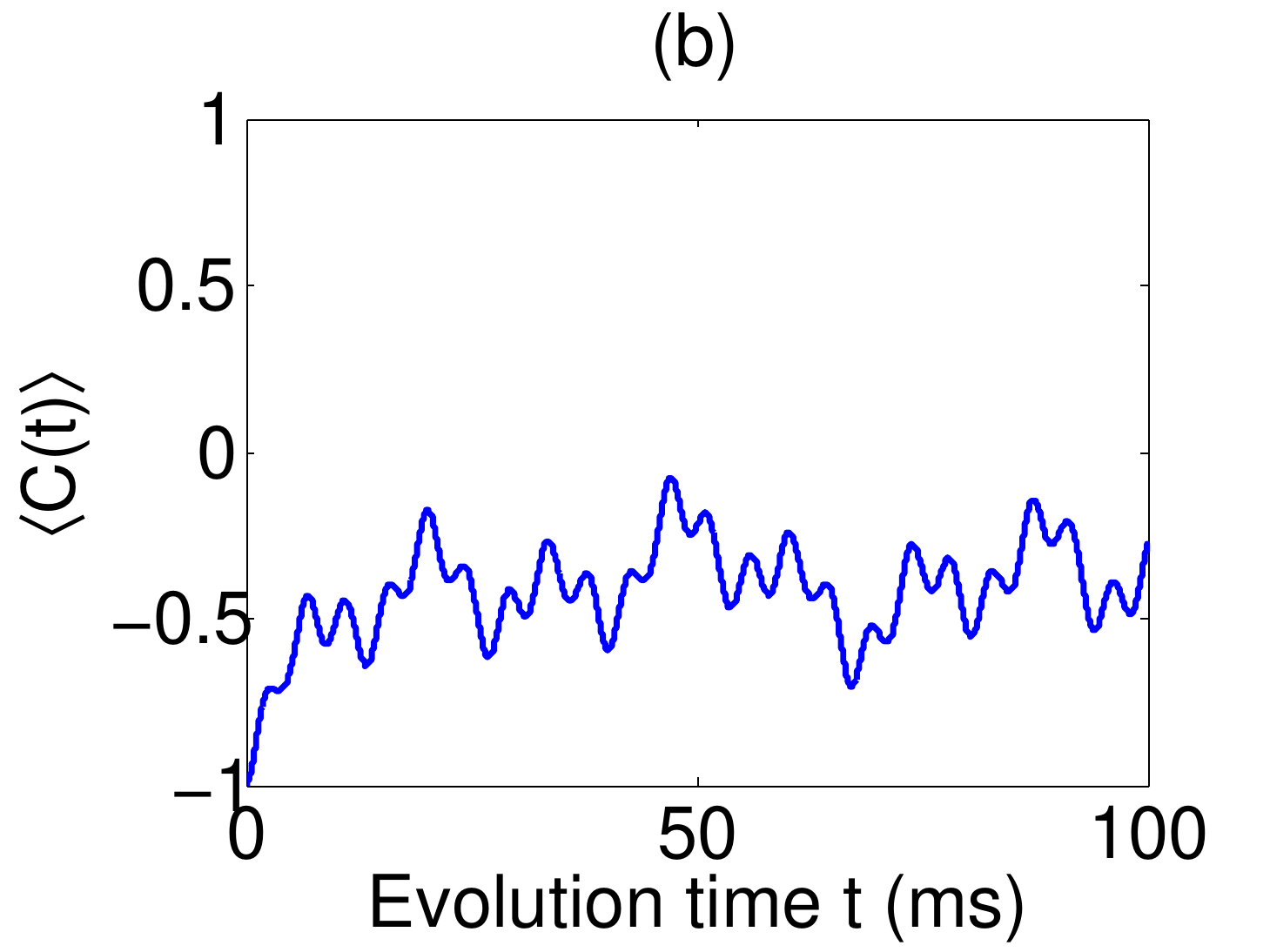}

\includegraphics[width=0.25\textwidth]{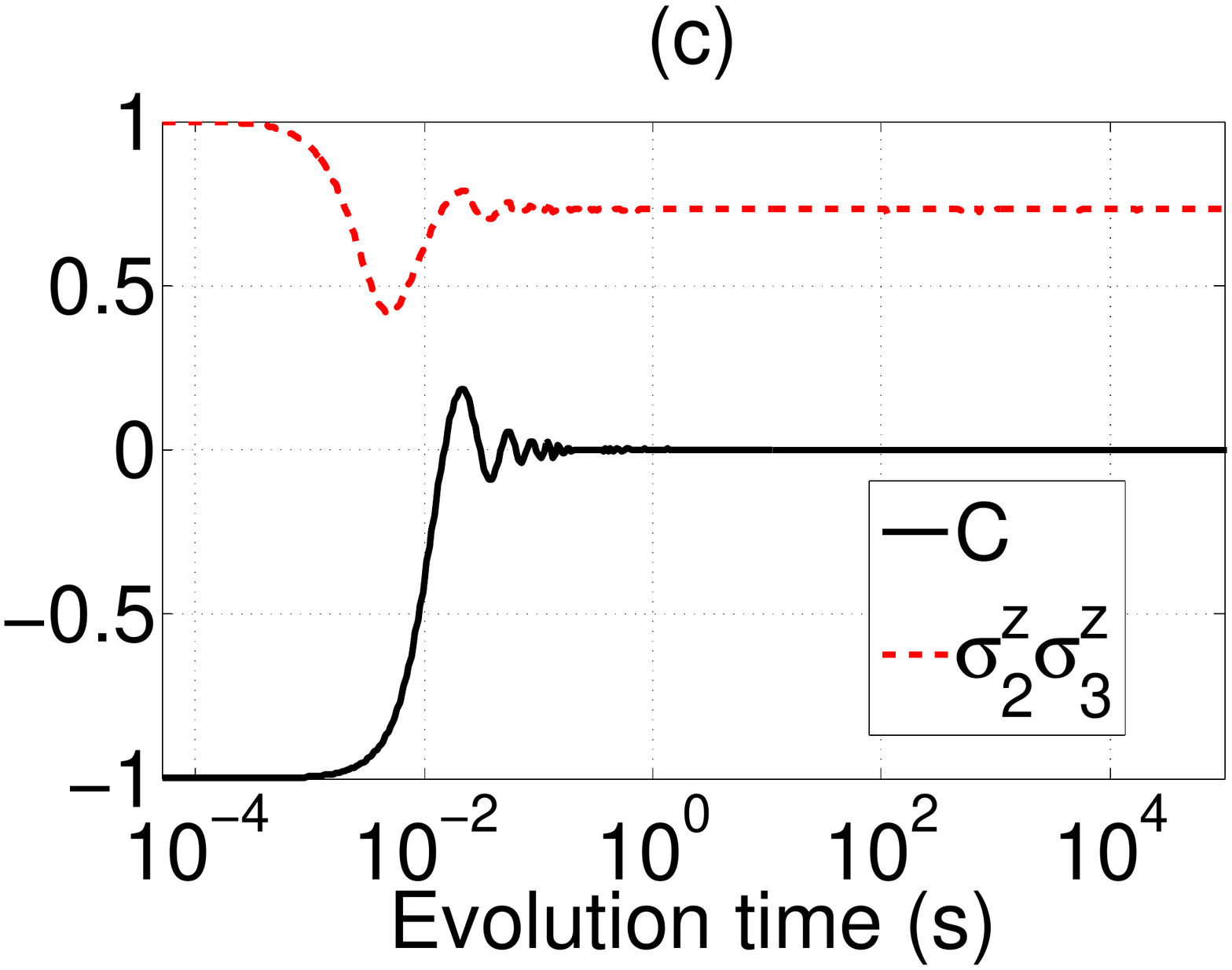}\includegraphics[width=0.25\textwidth]{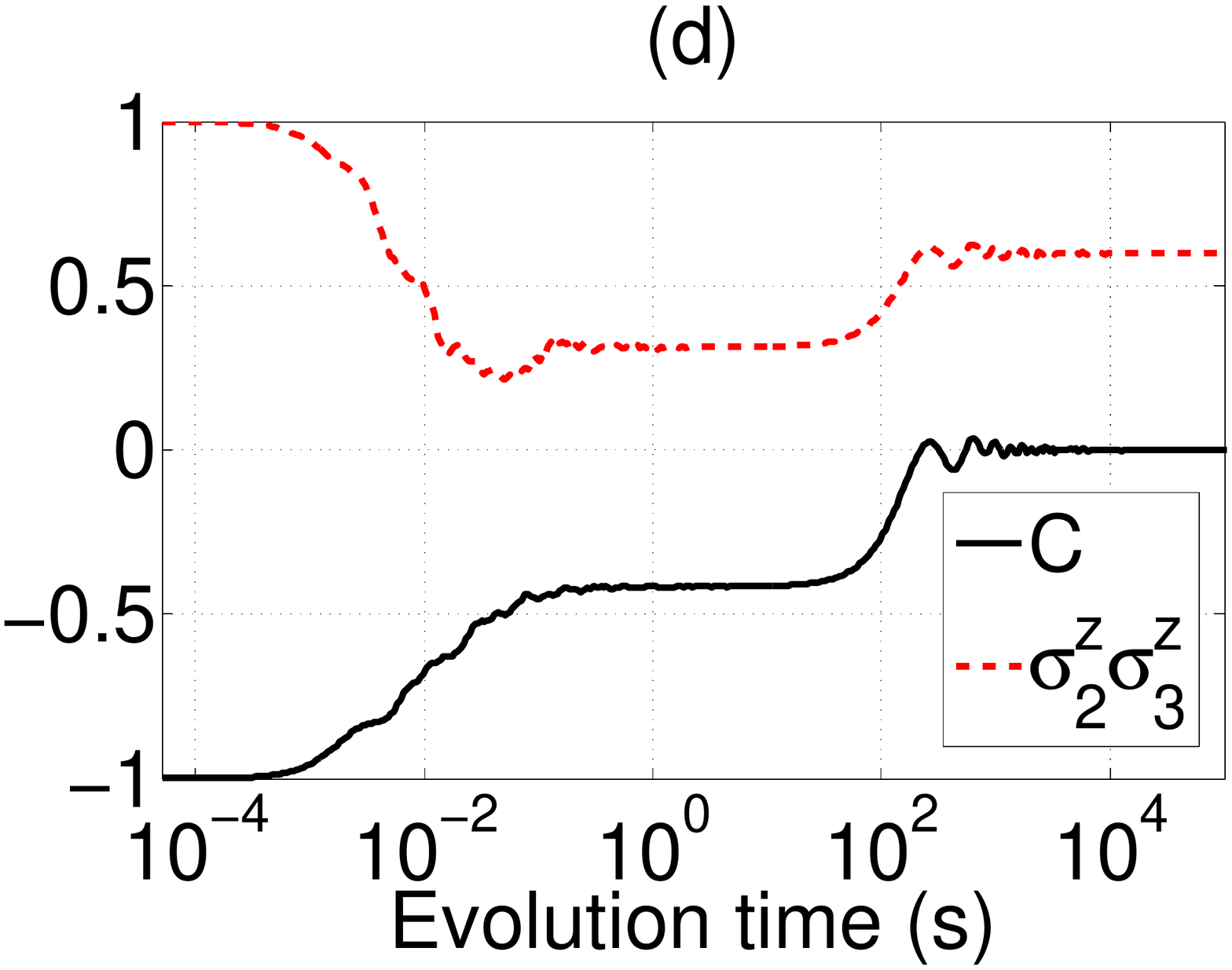}

\caption{\label{fig3}(a-b) Short time dynamics of $\sigma_{i}^{z}$ and $C$
for short-range (a) and long-range (b) interaction. (c-d) Long time
dynamics of time-averaged $C$ and $\sigma_{i}^{z}\sigma_{j}^{z}$
for (c) short-range and (d) long-range interaction. The parameters
for short-range and long-range interaction are given by Fig. 2}
\end{figure}

For these choices of parameters, the short time dynamics with $t\in[0,2/J_{0}]$
for all $\langle\sigma_{i}^{z}\rangle$ and $\langle C\rangle$ are
shown in Fig. \ref{fig3}. In the short-range interaction case, one
observes that the spin excitation, initially located at the left edge
of the chain, almost coherently travels to the other side and oscillates
back and forth with relatively small dispersion. In contrast, the
spin excitation diffuses to the rest of the chain in the long-range
interaction case and somehow get locked before it reaches the middle
of the chain (with $\langle C\rangle\approx-0.4$).

\begin{figure}
\includegraphics[width=0.4\textwidth]{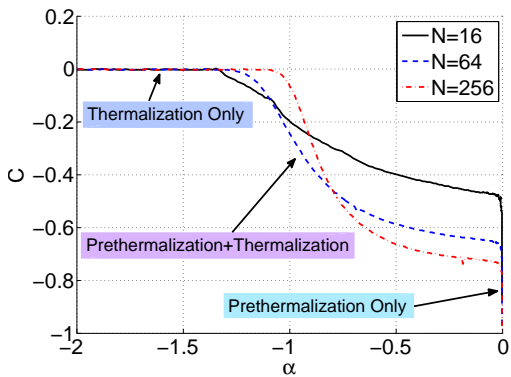}

\caption{\label{fig4} A dynamical ``phase diagram'' with regard to the interaction
range characterized by the fitting parameter $\alpha$. The $N=16$
case uses the parameters specified in Fig. \ref{fig1} and $N=64,256$
cases use the same parameters except that $\omega_{z}$ is scaled
down by $\omega_{z}\propto\sqrt{\ln N}/N$ to maintain chain stability.}
\end{figure}

To better show the long time dynamics, we use the finite-time-averaged
quantity $\overline{A(t)}$, defined as $\overline{A(t)}\equiv\frac{1}{t}\int_{0}^{t}\langle A(\tau)\rangle d\tau$
for the observable $A$ \cite{rigol_hard-core_2006,ponomarev_thermal_2011}.
This will average out temporal fluctuations on short time scale (the
following dynamical behaviors are qualitatively the same even without
performing any time averaging). The long time dynamics is shown in
Fig. \ref{fig3}(c-d). In the short-range interaction case, the spin
excitation position $C$, as well as the spin correlation $\sigma_{i}^{z}\sigma_{j}^{z}$,
relax to the stationary values at around $T_{0}\approx10/J_{0}=500ms$.
In the long-range interaction case, observables first reach quasi-stationary
(prethermal) values at time scale $T_{0}$, and further relax to the
stationary (thermal) values at much longer time scale ($10^{4}T_{0}$).
The emergence of prethermalization is manifested in the long-range
interaction case by a nonzero value of $\overline{C}$ at intermediate
time scale $T_{0}$. We can use $\overline{C}$ as an order parameter
to characterize different dynamical behaviors. By continuously tuning
the effective interaction range (indicated by the fitting parameter$\alpha$)
with the beatnote frequency $\mu$, we find that (see Fig. \ref{fig4})
prethermalization only takes place when $\alpha$ is smaller than
a critical value $(\alpha_{C}\approx1.3$ for $N=16$). For larger
system size, the prethermalization-thermalization transition still
occurs, but $\alpha_{C}$ becomes smaller and the transition becomes
sharper. For the particular case $\alpha=0$, the system has uniform
coupling and its dynamics can be solved exactly. The exact solution
shows that the system stays in the prethermal state forever with $\overline{C}=\frac{2}{N}-1$.

\begin{figure}
\includegraphics[width=0.25\textwidth]{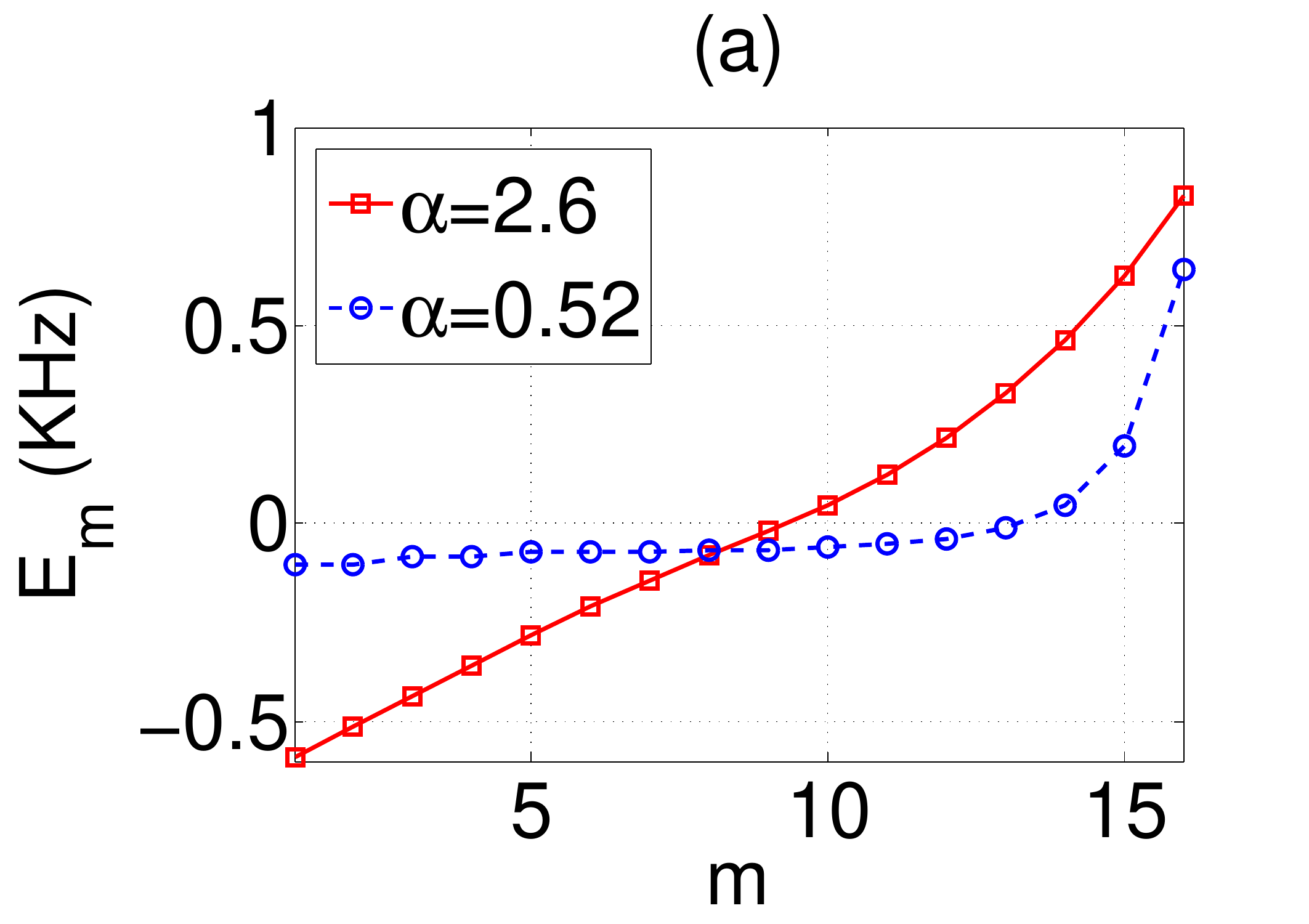}\includegraphics[width=0.25\textwidth]{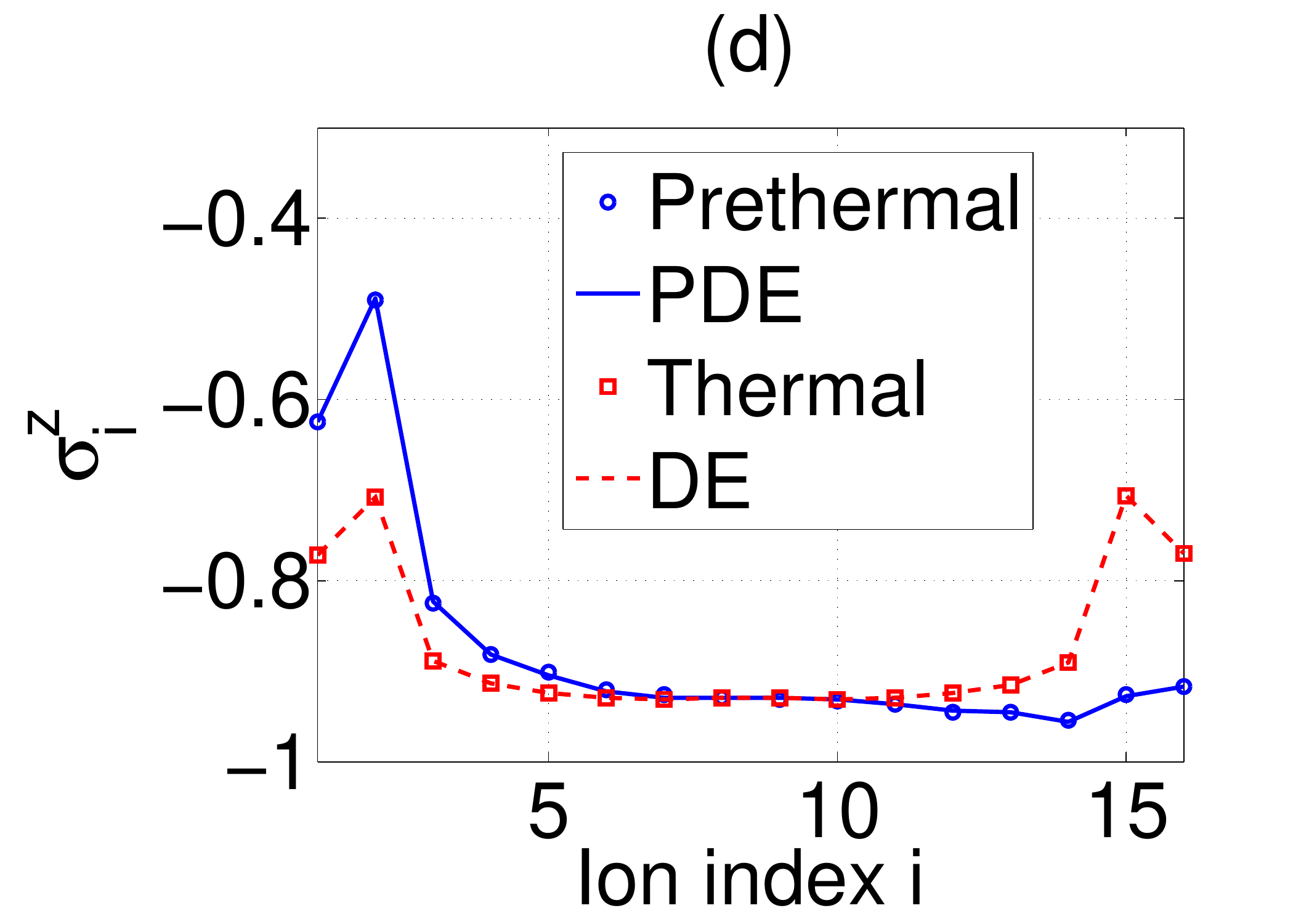}

\includegraphics[width=0.25\textwidth]{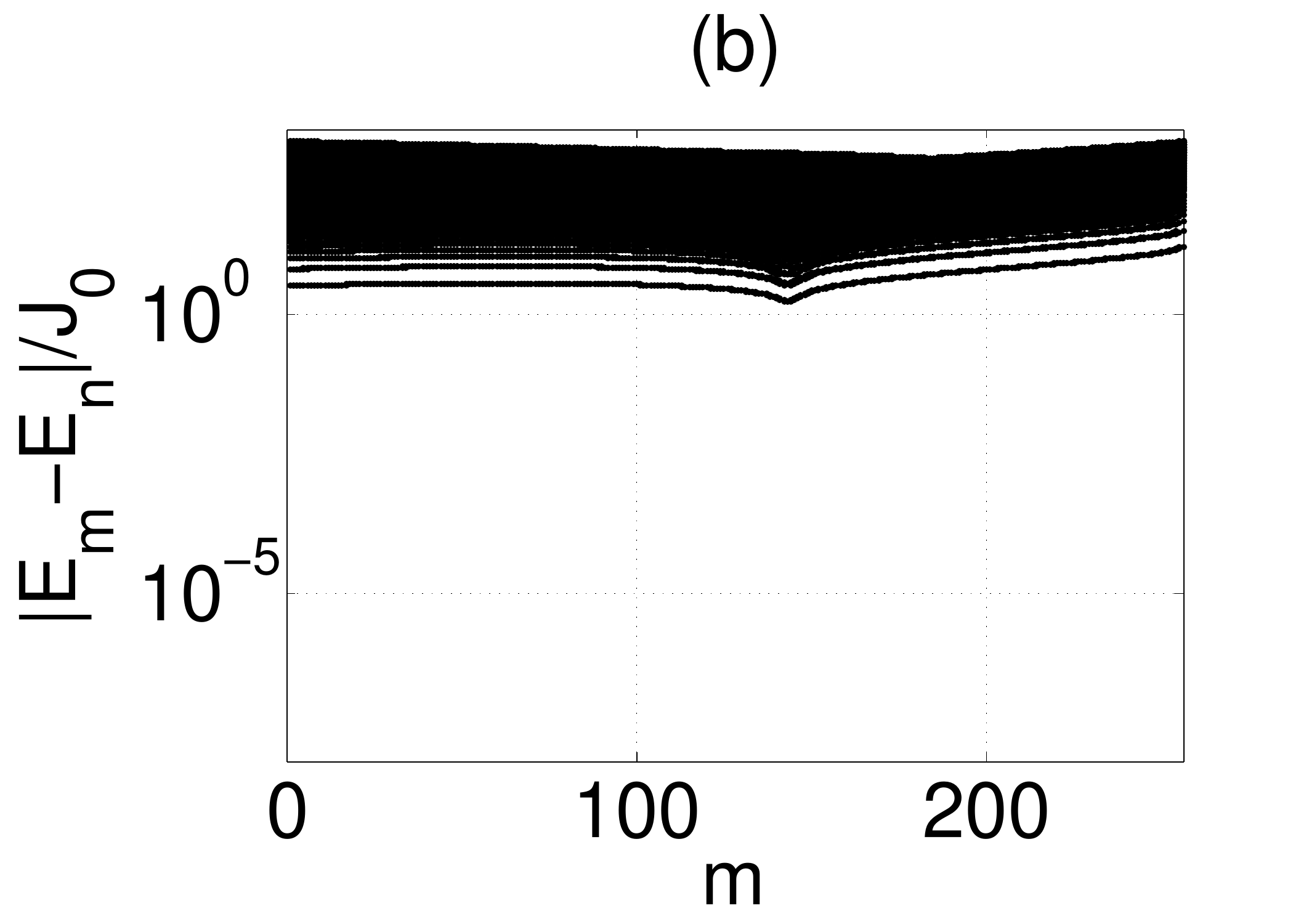}\includegraphics[width=0.25\textwidth]{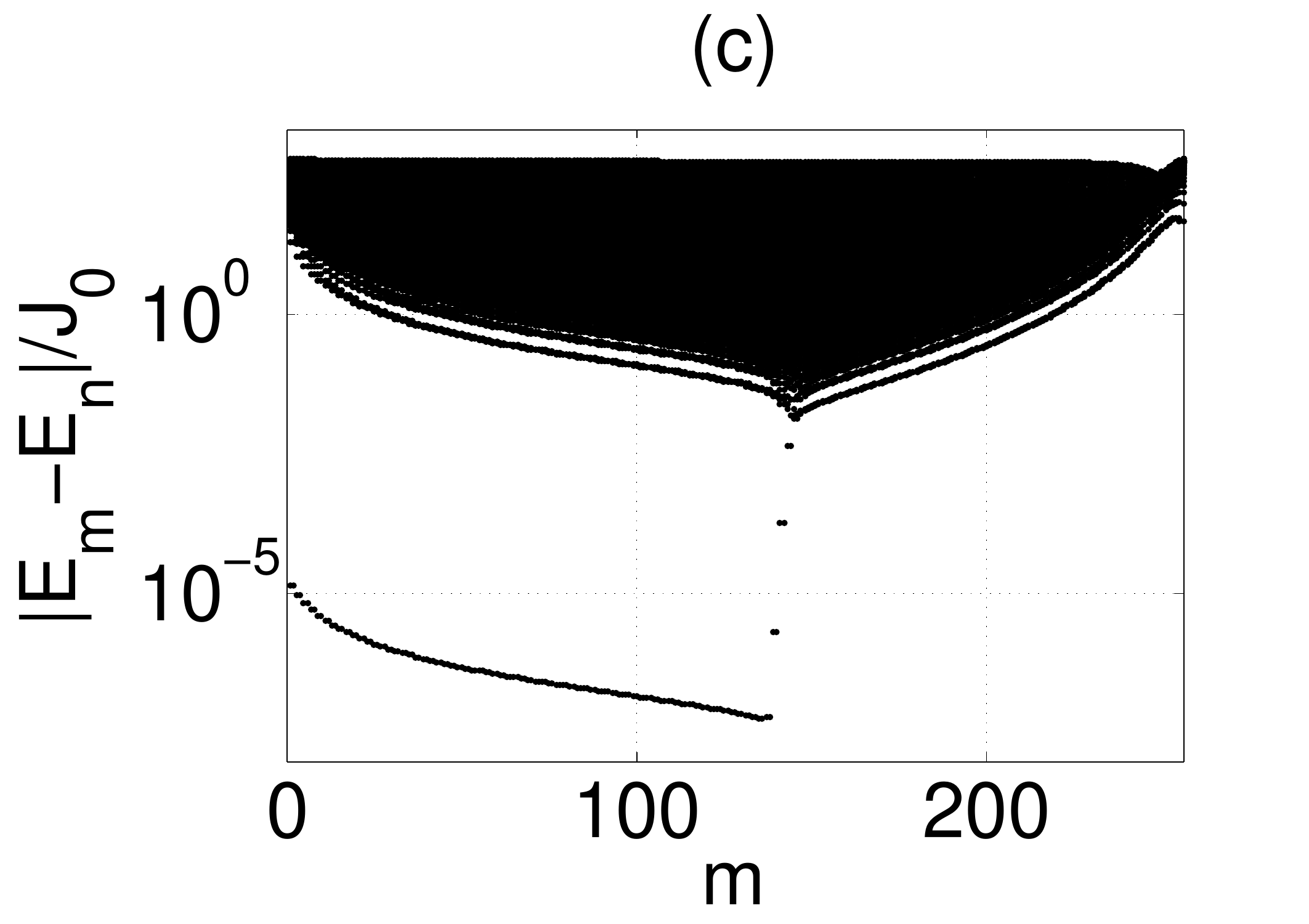}

\caption{\label{fig5}(a) Energy spectrum in the single spin excitation subspace
for $N=16$ spins with short and long range interaction. (b-c) Scatter
plot of eigenenergy differences $\{E_{m}-E_{n}\}$ for $N=256$ spins
with (b) short range interaction ($\alpha=2.4$) and (c) long range
interaction ($\alpha=0.74$). (d) Comparison of prethermal values
of $\overline{\sigma_{i}^{z}}$ (blue circle, taken at $t=T_{0}=500ms$)
with PDE prediction (blue solid line) and thermal values (red square,
taken at $t=5\times10^{3}s$) with DE prediction (red dashed line)
based on long range interaction pattern $J_{ij}$ shown in Fig. \ref{fig2}b.
See in the text for the definition of PDE and DE.}
\end{figure}

\emph{Mechanism of prethermalization and dynamical transition} . We
now give a physical explanation to why prethermalization and different
dynamical behaviors occur in this model. The distinctive short time
dynamics of $\langle\sigma_{i}^{z}(t)\rangle$ (Fig. \ref{fig3}a,b)
can be explained by examining the energy spectrum of $H_{XY}$ in
the single spin excitation subspace (shown in Fig. \ref{fig5}a).
In the short-range interaction case, the energy spectrum is close
to linear. This is because $H_{XY}$ can be roughly approximated with
only neighboring interaction, which is then identical to a tight-binding
fermionic model 
\[
H_{tb}=2\sum_{i}(J_{i,i+1}c_{i}^{\dagger}c_{i+1}+Bc_{i}^{\dagger}c_{i})
\]
and shows ``quantum mirror'' behavior \cite{albanese_mirror_2004,yao_quantum_2013},
resulting in a near dispersion-free spin wave propagation until non-linearity
sets in. On the other hand, the energy spectrum for the long range
interaction case is highly non-linear, so the dynamics of spin excitation
is strongly dispersive.

The cause for existence of prethermal stage in the long time dynamics,
however, is much more complicated. Naively, the spin flip-flop matrix
$J_{ij}$ varies smoothly among sites for any $\alpha\in(0,3)$, so
the spin excitation should continuously diffuse from one end of the
chain to the whole chain, and is not expected to get trapped somewhere
in the middle for a long time. The two stage dynamics indicates that
there are two different time scales weaved in the Hamiltonian, which
is not at all obvious by looking at $J_{ij}$. We note, however, that
the time dynamics of any physical observable is simply given by 
\[
\langle A(t)\rangle=\sum_{m,n}\rho_{mn}(0)A_{nm}e^{i(E_{m}-E_{n})t/\hbar},
\]
where $\rho_{mn}(0)$ is the initial state's density matrix element
in energy basis. So different time scales of dynamics can be unraveled
through mapping of eigenenergy differences $\{E_{m}-E_{n}\}$, as
done in Fig. \ref{fig5}. In the short-range interaction case (Fig.
\ref{fig5}b), all $\{E_{m}-E_{n}\}$ are continuously distributed
from $J_{0}$ to $100J_{0}$, so a single-stage relaxation is expected
after $T\sim T_{0}=10/J_{0}$. In the long-range interaction case
(Fig. \ref{fig5}c), most $\{E_{m}-E_{n}\}$ still fall into the range
of $1-100J_{0}$, but there is a striking separate branch gaped at
much lower rate $(\sim10^{-6}J_{0}$). This branch actually corresponds
to near-degenerate pairs ($\{E_{2k}-E_{2k-1}\}$) of eigenenergy (Fig.
\ref{fig5}a) that make up the first half energy spectrum, and the
number of these pairs scale up with system size $N$. The appearance
of near-degenerate pairs $\{E_{2k}-E_{2k-1}\}$) in eigenspectrum
of our model seems to be due to a combined effect of long-range interaction
and inhomogeneous lattice spacing. If we put the ions into a ring
(or flat-bottomed) trap so that the ions are equally spaced, we find
that there is no separate branch in $\{E_{m}-E_{n}\}$ plot (figure
not shown) and hence no prethermalization behavior even with the same
long-range interaction. 

Figure \ref{fig5}d shows that the thermal values can be well predicted
by the diagonal ensemble (DE), defined as

\[
\rho_{DE}=\rho_{mn}(0)\delta_{mn},
\]
In large $N$ limit, under certain conditions prescribed as \emph{eigenstate
thermalization hypothesis} \cite{deutsch_quantum_1991,srednicki_chaos_1994,rigol_thermalization_2008},
the diagonal ensemble prediction will match the canonical ensemble
for thermalized state in classical statistical physics.

As prethermalization is due to a different scale of eigenenergy difference
in the Hamiltonian, to predict the prethermal values here, we can
define the \emph{partial diagonal ensemble} (PDE): 
\[
\rho_{PDE}=\begin{cases}
\rho_{mn}(0)\delta_{mn} & |\nu_{m}-\nu_{n}|\gtrsim1/T_{0}\\
\rho_{mn}(0) & |\nu_{m}-\nu_{n}|\ll1/T_{0}
\end{cases}
\]
where $\{\nu_{m}\}$ are the eigenenergies. We find that the PDE can
well predict prethermal values of local observables $\overline{\sigma_{i}^{z}}$,
as shown in Fig. \ref{fig5}d. Roughly speaking, the prethermalization
time scale is determined by the average level spacing ($\sim10/J_{0}$),
and the thermalization time scale is determined by the minimum level
spacing ($\sim10^{4}/J_{0}$ as in Fig. \ref{fig5}d).

The dynamical transition can be associated with breaking of the lattice
inversion (parity) symmetry, reminiscent of symmetry breaking in equilibrium
phase transitions. Our Hamiltonian $H_{XY}$ is symmetric under the
space inversion around $z=0$ (Fig. \ref{fig1}), but we start from
an initial state that does not have this symmetry. The thermal state,
with no memory of initial state, is described by the diagonal ensemble
$\rho_{DE}$ and restores this symmetry as $\overline{C}=0$. However,
the prethermal state does not restore the Hamiltonian symmetry due
to its non-zero $\overline{C}$ value, which indicates that some ``memory''
of initial state is preserved in prethermal state. The intermediate
time scale $T_{0}$ for observation of the prethermal state gives
a microscopic interpretation why this state can break the parity symmetry:
one cannot distinguish the near-degenerate pairs of eigenstates in
the energy spectrum, so linear combinations within each pair are allowed.
Since the two eigenstates of the pair have either even or odd parity,
their linear combinations can break the parity symmetry. The dynamical
phase diagram shown in Fig. \ref{fig4}) also has the hint of two
non-analytic points: one is where $\overline{C}$ becomes non-zero,
representing the appearance of prethermalization, and the other is
where $\overline{C}$ approaches $\frac{2}{N}-1$ ($\alpha\rightarrow0$),
representing the disappearance of thermalization. 

\emph{Discussion of experimental detection}: The transverse field
Ising Hamiltonian Eq. \ref{TFIM} has already been experimentally
simulated in Ref. \cite{islam_emergence_2013} for $N=16$ ions, with
demonstrated highly-efficient in-situ measurement of spin polarization
($\sigma_{i}^{z}$) and spin correlation ($\sigma_{i}^{z}\sigma_{j}^{z}$).
The XY Hamiltonian (Eq. \ref{XY}) can thus be readily obtained by
tuning up the effective transverse magnetic field. The non-equilibrium
initial state preparation requires focused laser beam, but is relatively
easy due to large ion spacing near the ends. The laser power and trap
frequencies used for generating interaction pattern $J_{ij}$ as shown
in Fig. \ref{fig2} are within current experimental reach \cite{islam_onset_2011,islam_emergence_2013}.
The observation of prethermalization and the dynamical transition
shown in Fig. \ref{fig3} \& \ref{fig4} only requires the spin decoherence
time longer than $T_{0}=10/J_{0}=500ms$, and coherence time up to
$2.5$s has been experimentally achieved using the hyperfine qubit
of $Yb^{+}$ ions \cite{olmschenk_manipulation_2007}. But the second
stage of thermalization for long-range interaction case will take
much longer time and is beyond current experimental reach, similar
to the experiment on prethermalization with cold atoms \cite{gring_relaxation_2012}.

In summary, we have proposed a novel scheme to observe the peculiar
prethermalization phenomenon and dynamical transitions in the experimental
system of trapped ion quantum simulator. The required conditions fit
well with the current experimental technology. We provide an explanation
of the mechanism of prethermalization and dynamical transition in
our proposed model, which is connected with some unique feature of
this experimental system.
\begin{acknowledgments}
We thank C. Monroe, A. Polkovnikov, and A. Gorshkov for helpful discussions.
This work was supported by the NBRPC (973 Program) 2011CBA00302, the
DARPA OLE program, the IARPA MUSIQC program, and the ARO/AFOSR MURI
program. 
\end{acknowledgments}
\bibliographystyle{apsrev4-1}
\bibliography{library}

\end{document}